# Essential L-Amino Acid-Functionalized Graphene Oxide for Liquid Crystalline Phase Formation

H. Gharagulyan[1, 2] *, Y. Melikyan[1], V. Hayrapetyan[1], Kh. Kirakosyan[1], D. A. Ghazaryan[3,4], M. Yeranosyan[1, 2]

[1] Innovation Center for Nanoscience and Technologies, A.B. Nalbandyan Institute of Chemical Physics NAS RA, 5/2 P. Sevak Str., Yerevan 0014, Armenia

[2] Institute of Physics, Yerevan State University, 1 A. Manoogian, Yerevan 0025, Armenia

[3] Center for Photonics and 2D Materials, Moscow Institute of Physics and Technology, Dolgoprudny 141701, Russia

[4] Laboratory of Advanced Functional Materials, Yerevan State University, 1 A. Manoogian, Yerevan 0025, Armenia

* Corresponding author at: A.B. Nalbandyan Institute of Chemical Physics NAS RA, 5/2 P. Sevak Str., Yerevan 0014, Armenia.
E-mail address: herminegharagulyan@ysu.am (H. Gharagulyan)

**Abstract**

**The colloidal 2D materials based on graphene and its modifications are of great interest when it comes to forming LC phases. These LC phases allow controlling the orientational order of colloidal particles, paving the way for the efficient processing of modified graphene with anisotropic properties. Here, we present the peculiarities of AA functionalization of GO, along with the formation of its LC phase and orientational behavior in an external magnetic field. We discuss the influence of pH on the GOLC, ultimately showing its pH-dependent behavior for GO-AA complexes. In addition, we observe different GO morphology changes due to the presence of AA functional groups, namely L-cysteine dimerization on the GO platform. The pH dependency of AA-functionalized LC phase of GO is examined for the first time. We believe that our studies will open new possibilities for applications in bionanotechnologies due to self-assembling properties of LCs and magnificent properties of GO.**

**Abbreviations:** GO: graphene oxide; rGO: reduced graphene oxide; LC: liquid crystal; GOLC: graphene oxide liquid crystal; AA: amino acids; His: histidine; Tyr: tyrosine; Cys: cysteine.

## Introduction

The dynamics of modern scientific and technological development imply that research into graphene-like materials will continue to be a priority for the scientific community. As a result, the findings of these studies can be directly applied to next-generation devices [1]. This, in turn, poses a challenge to explore and develop novel approaches or methods to obtain new derivatives of graphene, such as graphene oxide (GO) and reduced graphene oxide (rGO) [2]. The characteristic properties (mechanical, electronic, optical, and chemical) of devices based on these materials can be substantially enhanced and tuned through functionalization or surface modification [3]. Here, one of the intensively studied directions is the interaction of proteins and peptides with GO (oxygen-enriched version of graphene), in particular, the binding thermodynamics with various amino acids [3]. The interaction of GO with AAs is mainly determined by the electrostatics and van der Waals π–π interaction, which improves the electron transfer process. Due to the significant enthalpy–entropy correlation between GO and AAs, it can act as an artificial receptor for small molecules [5]. Furthermore, the noncovalent interaction of GO and AAs (GO-AA interaction) plays a major role in many applications: displays, transistors, self-powered devices, spatial phase modulators, saturation absorbers, switches, and sensors [6 - 8].

When anisotropic particles of sufficiently high concentration are suspended in liquid media, they can spontaneously demonstrate a liquid crystalline transition [9, 10]. The LC phase of GO was discovered because of processing by the wet chemical exfoliation method [11]. Three main factors affect its formation in GO suspensions: concentration, the average aspect ratio of flakes, and pH of solvent media [12]. Carboxylic and hydroxylic groups on the edges and the basal planes are charged functional groups, which can be protonated/deprotonated due to changes in pH [13]. The dispersed GO exhibits a stable nematic phase due to the repulsive electrostatic force generated by the hydrolysis of carboxylic-hydroxyl groups on the edges of its flakes. Notably, viscosity is another moderate factor with an important role in forming the LC phase. GO dispersions show a biphasic behavior (both isotropic and nematic phases coexist) at a concentration of ~ 0.25–0.75 mg/ml [13].

The self-assembling properties of LCs, along with the magnificent properties (mechanical, thermal, and electrical) of modified graphene, maximize the use of graphene-based materials [15, 16]. It is worth mentioning that the application of 2D material liquid crystals is possible via two well-known approaches: dispersion of 2D materials in a liquid crystalline host and the liquid crystal phase formation from dispersions of 2D material flakes in different solvents. Due to 2D materials' unique properties, their application possibilities are almost limitless; however, the exploitation of these materials in technology faces many challenges from the point of view of scalability, production cost, and device-tunability. While reconfigurability of the devices can be provided by the adoption of LC properties, the combination of exceptional physicochemical properties, in turn, provides the out-of-competition status of those highly ordered liquid crystalline 2D materials for diverse applications (development of films, fibers, and membranes, display applications, optoelectronic devices, and quality control of synthetic processes) [17]. In [18], the application of the self-assembling property and LC behavior of polymer-modified GO is discussed. Another interesting approach to the self-assembly guiding possibility of GO surfaces for optically active materials is reported in [19]. Thus, controllable GOLC alignments by external fields are in strong demand and still pose a challenge [20]. Synthesizing GOLC is challenging due to difficulties in obtaining uniform flake size and homogeneous distribution. The second complication is the provision and control of the high alignment regulation parameter of GO. However, it is essential to understand which graphene derivative is in demand, as they have fundamentally different properties. For instance, graphene is a semimetal with zero bandgap, while the bandgap in GO is ~ 2.2 eV. For rGO, it varies between ~ 1.00 and 1.69 eV depending on the reduction degree [21]. It is worth mentioning that deoxygenation processes fail to completely remove oxygen-containing functional groups from the graphene's surface [22]. The review paper [23] focused on recent advances in preparing new functional composites using GO, rGO, and bio-based macromolecules. The various rGO-$M_xO_y$ composite materials were synthesized by microwave irradiation, which are potential candidates for future field emission, sensing, and storage devices [24-26]. Previous studies [27, 28] summarized the most recent studies on the laser-assisted synthesis of graphene-based materials and their modifications for application in energy storage devices, such as supercapacitors and batteries as electrodes. In addition to composite materials, the optimized nanocomposite matrix was fabricated by three carbon nanomaterials (C-dots, SWCNTs, and rGO) in [29], which are promising in wastewater treatment. Different techniques for suspending particles from the suspension medium to a substrate are used: namely, the electrophoretic method for GO and rGO deposition described in [30].

GO contains functional groups, such as carboxyl, carbonyl, epoxy, and hydroxyl, capable of forming chemical bonds with the ligand (e.g., P, N, O, or S) [31]. On the one hand, its functionalization with organic functional groups allows modifying the electronic properties, for instance, a bandgap tuning [32]. On the other hand, the adsorption of amino acids by 2D materials considerably changes their biological response in terms of reactive oxygen species generation [33].

Fluorescence titration and isothermal titration calorimetry allow describing the formation of GO-AA complexes [4, 34]. The former is used to estimate the GO-aromatic AA interaction and provide information on molecular bindings and their binding thermodynamics and complexation. S. Pandit et al. described binding constants, Gibbs free energy changes, enthalpy changes, and entropy changes for the complexation of various AAs with GO [4]. They proposed that electrostatic and van der Waals π–π interactions independently and complementarily control the complexation process of GO-AA and play a significant role in interactions depending on an AA type. Notably, GO interacts strongly with aromatic amino acids through π–π stacking, charge transfer, and H-bindings [35]. Thus, knowing the interaction mechanisms of GO-AAs complexes, we can explore the full potential of their LC phases and extensively advance them as they are widely used in the development of devices for many fields, such as industrial, environmental, and biomedical research. This is due to their highly ordered structures combined with exceptional mechanical and conductive properties [36, 37]. This work explores the synthesis and characterization of AA-functionalized GO, rGO, and LC phases, their mass/volume fractions, and pH values. Magnetic properties of the GOLC-AA complexes were also examined.

## Materials and Methods

**Materials.** The pyrolytic graphite, the solvents, amino acids, and all the chemicals used in the experiments were purchased from Sigma-Aldrich Chemical Co.

**GO Synthesis.** GO was synthesized by an approach similar to that in [38, 39]. Graphite powder (0.75 g) was mixed to a 9:1 ratio in concentrated $H_3PO_4/H_2SO_4$ (10:90 mL) solution, where 4.5 g of $KMnO_4$ was later added. The obtained mixture was stirred at 50°C for 24 hours and allowed to cool at room temperature. Afterward, it was poured onto ice (200 mL). Before that, 3 mL of 30% $H_2O_2$ was added. The unpurified and crude product was centrifuged (4000 rpm, 30 min), allowing the supernatant to be decanted. Subsequently, the product was washed several times with water (400 mL), 30 % HCl (400 mL), and ethanol (400 mL). Then, ether (400 mL) was added to aid the coagulation. The final suspension was collected by filtration and vacuum-dried for 24 hours at room temperature.

**rGO Synthesis.** rGO was synthesized by an approach similar to that in [40]. 100 mg of GO was poured into a 250 ml round bottom flask, and then 100 ml of water was added to acquire a yellowish-brown dispersion. An ultrasonication was performed until a homogeneous solution was obtained. Next, 1.0 ml of hydrazine hydrate (32.1 mmol) was added to the solution and heated in an oil bath at 100°C for 24 hours by reverse distillation. Here, the rGO was gradually released as a solid precipitate. The resulting mass was filtered and washed thoroughly with deionized water (5 x 100 ml) and methanol (5 x 100 ml). After all these steps, drying was done under constant airflow.

**Synthesis of GOLCs.** LC dispersion of GO was prepared using centrifugation-induced size-fractionation. GOLC was formed by proper selection of the solvent, average aspect ratio of GO flakes, and the correct control of their concentration, i.e., via the dispersion of 0.5–5 mg/mL GO in water and ethanol. The exfoliated GO flakes were first filtered and dried to obtain the powder. Next, they were mixed and stirred with the solvent at room temperature. The colloidal nature of the solution was confirmed by the Tyndall effect [41].

**Preparation of AA-functionalized GO.** Aromatic AAs, such as L-tyrosine (Tyr-O donor), and nonaromatic AAs, such as L-histidine (His-N donor) and L-cysteine (Cys-S donor), were used to prepare solutions in deionized water at room temperature. The AA concentrations were 0.014 g for Tyr, His, and Cys in 200 ml DI water. Then, 0.025g/10mL of GO/ethanol solution was prepared. For all the AAs, neutral, basic, and acidic mixtures at different pHs were prepared (Table 1). The pH was adjusted with NaOH and HCl solutions. AA functionalization of GO was done following the approaches reported in [35, 42].

| Solution | Initial pH | pH after adding HCl | | pH after adding NaOH | |
|---|---|---|---|---|---|
| L-Tyr | 6.24 | 3.03 | 4.94 | 8.37 | 10.03 |
| L-His | 7.12 | 3.07 | 5.07 | 8.01 | - |
| L-Cys | 4.64 | - | - | 7 | 8.06 |

Table 1. pHs of prepared AA solutions.

## Characterization

The crystallographic data of GO and rGO powders were obtained by XRD (MiniFlex, Rigaku). The chemical compound analysis of GO and rGO was performed by FTIR Spectrometer (Spectrum Two, PerkinElmer). Raman spectroscope (XploRA PLUS, HORIBA) was used to identify the bond types/hybridizations of GO and rGO. The purity of GO was checked by ICP-MS (ELAN-9000, PerkinElmer). The mass loss of GO and rGO was measured over time using TGA (TGA 8000, PerkinElmer). Moreover, the optical characterization of GO and rGO was obtained by PL spectroscope (Cary Eclipse Fluorescence Spectrometer, Agilent). The absorbing properties of dispersed GO and rGO were investigated by UV-Vis spectrophotometer (Cary 60, Agilent). GO's liquid crystalline phase formation was observed with a polarizing microscope equipped with a digital camera (MT9930L, MEIJI Techno). The GO morphology and chemical composition was analyzed by SEM (Prisma E, Thermo Fisher Scientific). AFM characterization was done to identify the thickness of the synthesized material (NT-MDT Next).

## Results and Discussion

### Structural Analysis of GO and rGO

The characteristic spectra of the synthesized GO and rGO are presented in Fig. 1. To confirm the structure of the synthesized GO, the UV-Vis spectra of the aqueous solution of GO were recorded (see Fig. 1a). According to the absorbance spectra, the two peaks are due to π–π * and n–π * transitions; namely, the first peak of GO absorption is at 235 nm due to the transition of the aromatic C=C ring and the second at 310 nm due to transition of C=O bond. In the case of rGO, due to an increase in aromatic rings and a decrease in oxygen functional groups, the absorption peak is red-shifted to 260 nm (the restoration of the conjugated structure).

Fig. 1b shows the photoluminescence (PL) spectra of the aqueous solution of GO. The spectrum is obtained by changing the wavelength of the laser excitation wave from 380 to 396 nm by 4 nm. As can be seen from the spectra, the peak of photoluminescence was obtained at 402 nm. PL emission peak is decreased and blue-shifted to 398 nm after the GO reduction process, which is due to the $sp^2$ clusters expansion and decrease of the oxide functionalized groups.

The infrared absorption spectra of GO and rGO are presented in Fig. 1c. The GO spectrum shows the C=O carbonyl stretching at 1741 $cm^{-1}$, C=C at 1623 $cm^{-1}$, the C-O epoxy group stretching at 1219 $cm^{-1}$, and the last supplementary peak observed at 1042 $cm^{-1}$, which matched with C-O (alkoxy) group, respectively. The high intensity of the main peaks in the spectra confirms the presence of a large amount of oxygen functional groups after the oxidation process. From all this, we can deduce that we really managed to get GO. After the GO was reduced, all the functional groups were significantly decreased; the phenol C=C ring stretching at 1623 $cm^{-1}$ was present. For the rGO, the carboxyl group is absent on the spectrum.

The thermal stability of the synthesized GO was examined by thermogravimetric analysis. Heating was performed at 10°C per minute at a flow rate of 60 ml/min. As shown in Fig. 1d, in the case of the synthesized GO, in contrast to graphite which exhibits one clear step of weight loss, GO decomposes in two steps. In the first stage, the mass loss occurs at 50°C to 100°C due to the loss of water molecules; in the second stage, it occurs at 210°C –250°C due to the loss of oxygen-containing groups. The synthesized rGO shows a similar characteristic but with a lower amount of weight loss than that of GO due to a smaller amount of oxygen functional groups in the structure.

The XRD spectrum of synthesized GO powder is shown in Fig. 1e. Compared with pristine graphite, which has a diffraction peak at 26.4°, the peak of chemically exfoliated GO is shifted to lower angles having a maximum at 10.9°, and rGO exhibits a peak at 23.5°. These shifts are due to different interlayer distances between carbon base planes due to the intercalation process of oxygen functional groups.

The Raman spectra of GO and rGO are presented in Fig. 1f. Two sharp, intense peaks at 1345 and 1603 cm$^{-1}$ represent the D and G bands of the GO, revealing that the value of the $I_D/I_G$ ratio is 0.863, indicating the rarity of defects. A less intense 2D peak for GO is seen at 2682 cm$^{-1}$. The 2D band in GO was suppressed due to oxidation. For the rGO, the corresponding peaks' ratio is 0.916.

The thickness and morphology of the synthesized GO were determined via AFM analysis shown in Fig. 2. Then, 10 μm flake was observed, some part of which was monolayer and the other part bilayer due to folding at the edges. The thickness of the single-layer GO was ≈1 nm.

GO flakes, after reduction, were assembled on the sapphire substrate through dip coating (see Fig. 3a). Fig. 3b represents the optical microscope picture of the GO monolayer obtained by the drop-casting technique.

Homogeneous GO and rGO thin films were prepared by a CY-SP8-UV spin coater (4000 rpm, vacuum flow of at least 15 l/min). The morphology of the samples was studied using a scanning electron microscope (SEM) (see Fig. 3c). From the optical microscope images, we are finally convinced that we obtained a large-scale ~ 10 × 28 μm GO monolayer (GO was drop-casted on Si/SiO$_2$ (300 nm) substrate).

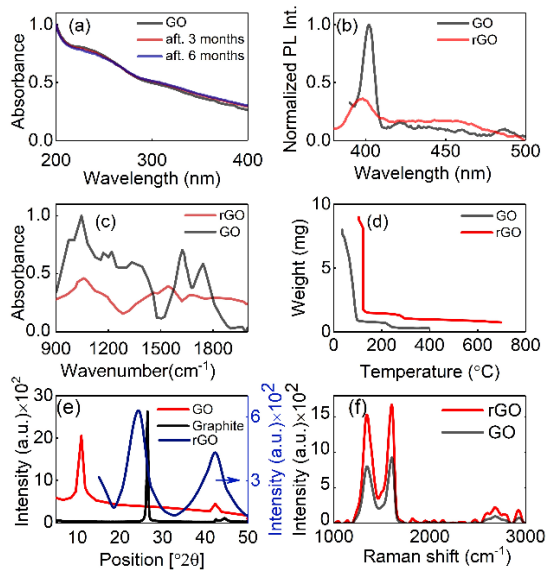

Fig. 1. Initial characterization of GO and rGO: (a) UV-Vis spectra of GO. Black line corresponds to the first day of measurements, (b) photoluminescence spectra of GO and rGO in aqueous solution, (c) FTIR-ATR spectra of GO and rGO, (d) thermogravimetric analysis of GO and rGO, (e) XRD analysis of graphite, GO, and rGO, and (f) Raman spectra of GO and rGO.

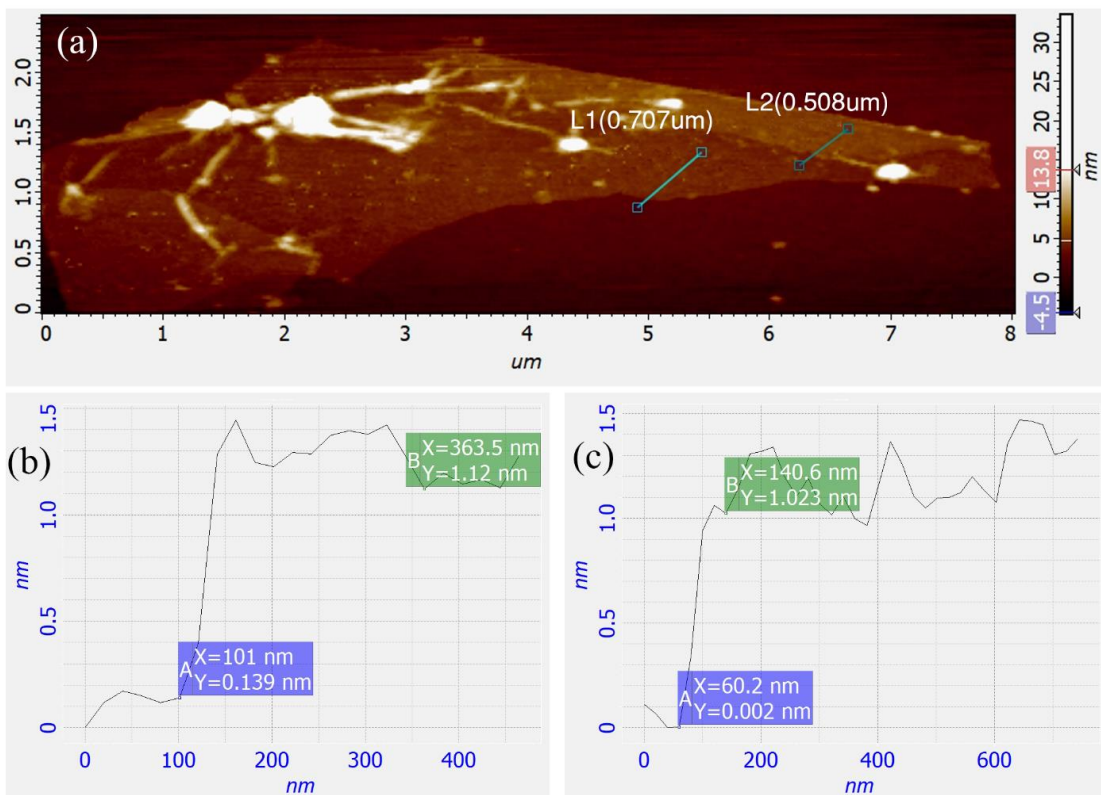

Fig. 2 a) AFM image of GO monolayer on the Si/SiO$_2$ wafer by drop-casting. Height profiles along b) L1 and c) L2 lines.

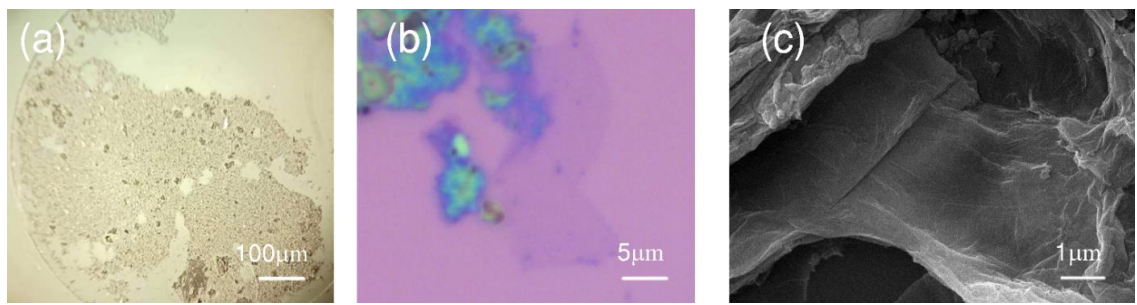

Fig 3. Optical micrographs of a) dip-coated rGO flakes, b) GO monolayer, and c) SEM image of GO.

Inductively coupled plasma mass spectrometer (ICP-MS) studies show less than 0.1 % of side product detection generated from the reaction during the synthesis. The GOLC phase was studied using a polarized optical microscope (POM) in transmission mode. The evolution from the isotropic phase to the LC phase was observed along with the gradual increase of GO concentration in the solvent (Fig.4). Phase transition is accompanied by the self-assembly or self-stacking of GO flakes. As can be seen from the POM images, the GOLC cell between the crossed polarizers gives a bright picture, demonstrating the existence of the birefringence.

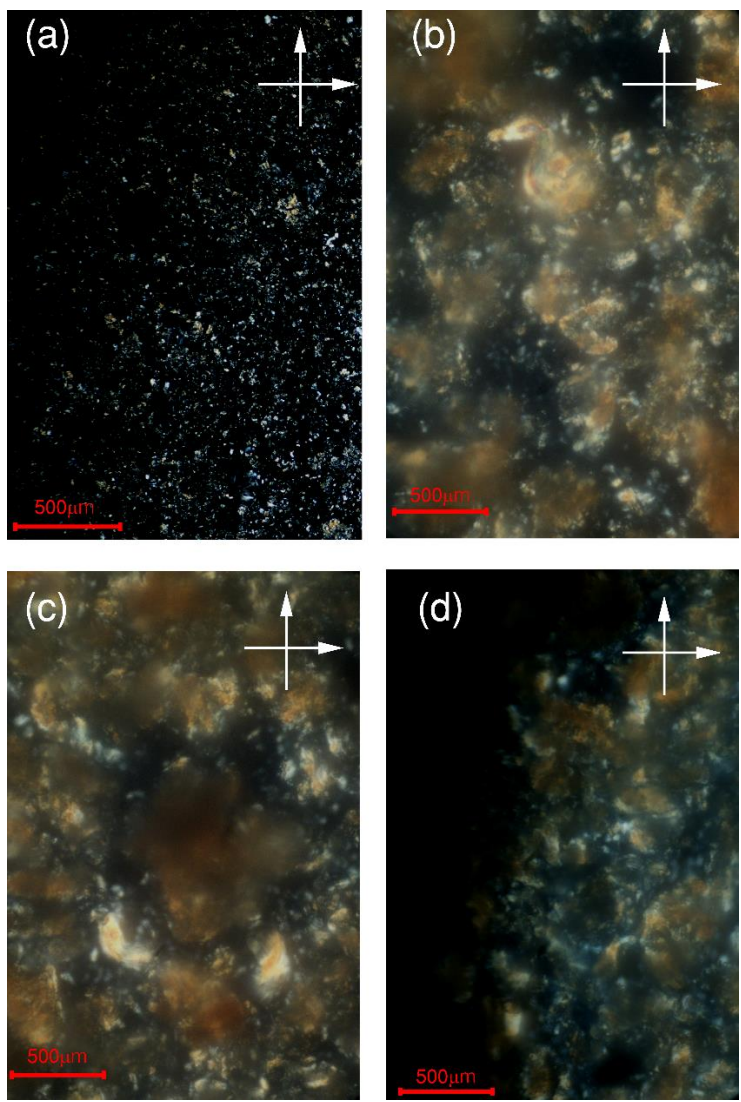

Fig. 4. Phase diagrams of GO dispersion in ethanol as a function of concentration. Typical optical micrographs in (a) biphasic and (b, c, and d) LC phases between crossed polarizers and scanned from different parts of the cell.

The analysis of all the measurements showed that the characteristic spectra and microscopic pictures of the synthesized material and its liquid crystalline phase correspond to the known ones in the literature [10].

**Amino Acid-Functionalized GO**

The UV-Vis spectrometric titration method was used to characterize the possible ways of GO-AA complexation and the structural peculiarities of GO and AAs. Electrostatic and $\pi–\pi^*$ interactions are critical contributors to GO-AA interaction. First, the pH dependencies of the AAs were studied (see Fig. 4). Then, the functionalization of GO by AAs was examined. Fig. 5a represents the pH dependency of L-Tyr; a 20 nm redshift was observed in the basic environment due to the deprotonation of the carboxyl group. In the acidic environment, only an increase in the absorption coefficient was observed due to the protonation of the amino group. Histidine exhibits a basic character in an aqueous medium due to its N-donating nature. Changing from a basic to an acidic medium, we observed a slight decrease in the intensity of the absorption peak and a shift to the longer wavelength range. At low pH, L-His becomes positively charged, disrupting existing hydrogen bonds and leading to electrostatic repulsion (see Fig. 5b). Cysteine has strong acidic properties and shows reversed behavior; the absorption peaks shift toward short wavelengths changing from basic to the acidic solution (see Fig. 5c).

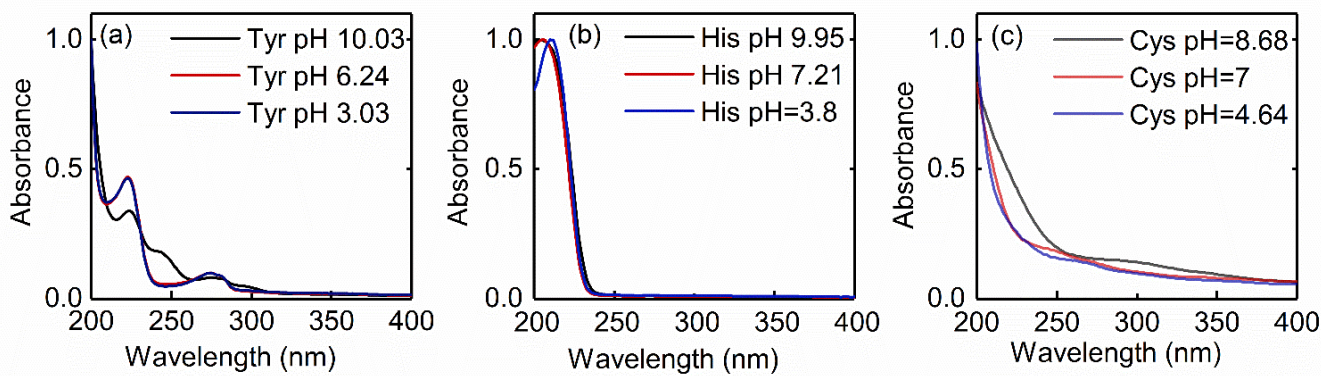

Fig. 5. UV-Vis absorption spectra of AAs at different pHs: (a) Tyr, (b) His, and (c) Cys.

First, the pH-dependent nature of GO-dispersed solutions was studied to reveal the pH dependency of AA-functionalized GO. Fig. 6 shows UV-Vis absorption spectra of GO-dispersed solutions for acidic, basic, and neutral pH values. As can be seen from the spectra, the absorption peaks are sharp at an acidic pH value.

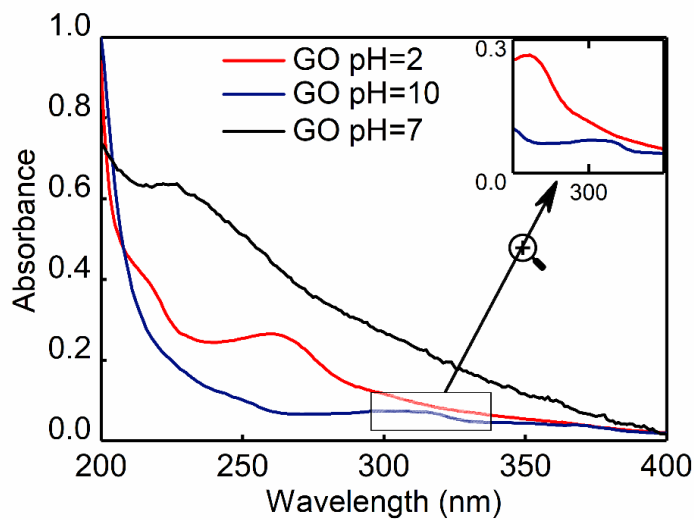

Fig. 6. UV-Vis absorption spectra of GO at acidic, basic, and neutral pH values.

**L-Tyr-Functionalized GO.** By gradually adding a certain amount of Tyr to the GO-dispersed ethanol solution, we extracted the UV-Vis spectra of the GO-Tyr system and observed the presence of an isosbestic point (the same extinction coefficient), implying that we have a clear complex formation (see Fig. 7). In addition, neither the large GO/rGO peak (260 nm) nor the Tyr peak (274 nm) was observed in the final GO-Tyr system. The redshift of the GO peak is due to the large size of the synthesized flakes and partial reduction.

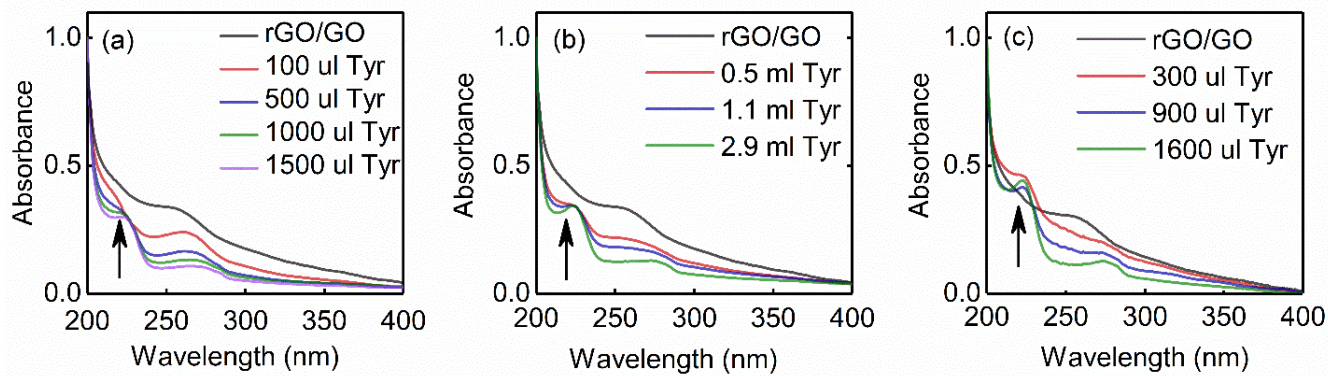

Fig. 7. UV-Vis absorption spectra of the GO-L-Tyr system at (a) pH = 3.06; (b) pH = 10.03; (c) pH = 6.24.

It was shown that in the case of aromatic AA, the interaction with GO is due to the π–π* stacking because the absorption peak corresponding to the 230 nm is increasing. The polar side chains and charge of amino acids are essential for GO-AA complexation. Polar amino acids, such as Tyr and His, form hydrogen bonds, while charged amino acid side chains form ionic bonds.

**L-His-Functionalized GO.** A similar study was performed for His, and again the presence of an isosbestic point was observed (Fig. 8). Neutral and basic GO-His systems show almost similar behavior, while for acidic solutions, absorption peaks are broadening due to the solvent-solute interactions' peculiarities. Polar solvents, such as water and ethanol, exhibit a stronger binding to solute through induced dipole-dipole interactions or hydrogen bonding.

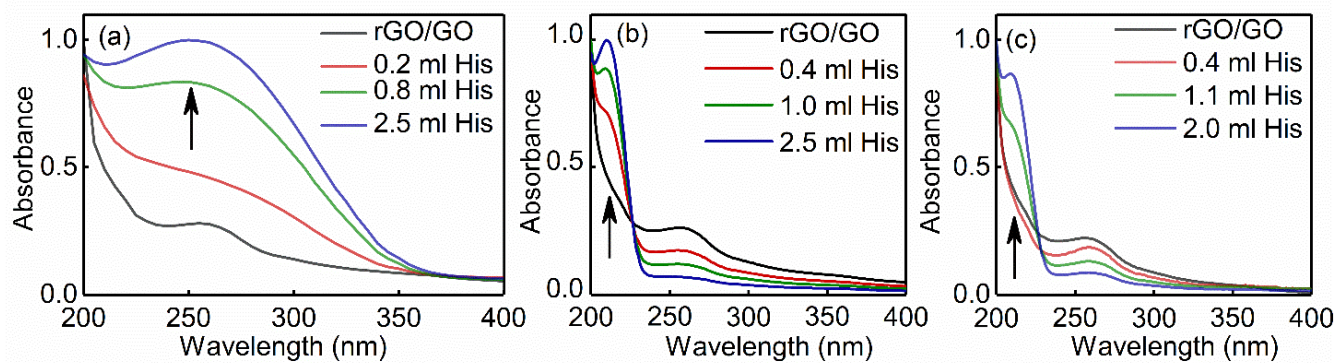

Fig. 8. UV-Vis absorption spectra of the GO-L-histidine system: (a) pH = 3.03; (b) pH = 10; (c) pH = 7.12.

**L-Cys-Functionalized GO.** Based on absorption spectroscopy, studies were also performed for the GO-Cys system at different pH values. Again, an isosbestic point was shown, confirming the successful implementation of the complexation (Fig. 9).

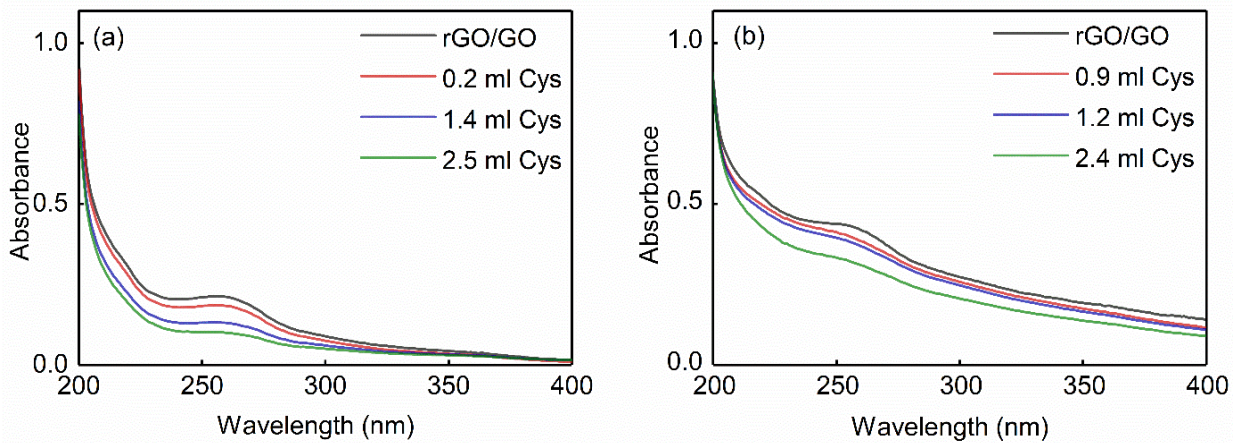

Fig. 9. UV-Vis absorption spectra of the GO-L-cysteine system: (a) pH = 4.64 and (b) = pH = 8.06.

From the obtained results by spectrometric titration, it can be concluded that the reason for complex formation is π−π* stacking and hydrogen bonding. As can be seen from the titration experiment, different amino acids interact to a different extent; namely, the GO complexation with AAs shows better results for an aromatic ring containing AAs.

XRD spectra of the functionalized GO are shown in Fig. 10. Compared with pristine GO, which has a diffraction peak at $10.9^0$, the peak of GO-Tyr is shifted to the lower angles having a maximum at $10.2^0$, and significant changes in intensity were recorded. GO-His exhibits a peak at $5.9^0$. An interesting behavior was observed for the Cys functionalized GO. In the XRD pattern of GO-Cys, the emphasized crystal structure was evident at $28.4^0$. As shown in [43], dimerization of L-Cys was feasible. As a result, L-cystine nanotubes were grown on the GO platform. This explanation was confirmed additionally by XRD, IR, Raman, and particle size analysis provided in the Supplementary Materials (see Fig. S1-Fig. S4).

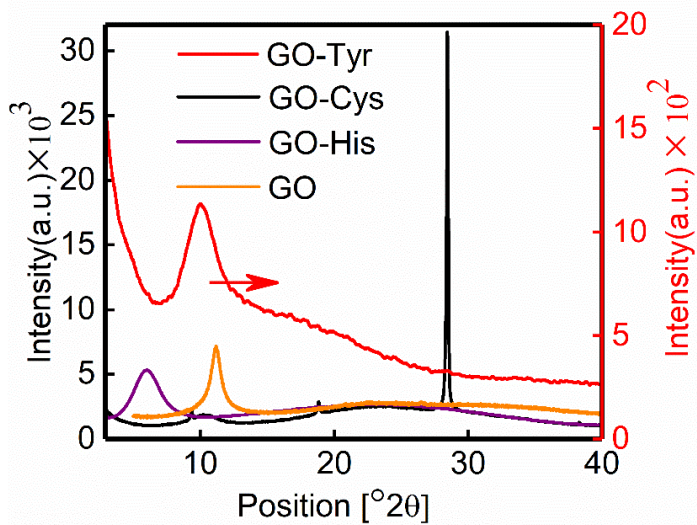

Fig. 10. XRD pattern of AA-functionalized GO.

Raman spectra of the GO for neutral, basic, and acidic pH values are shown in Fig. 11. The $I_D/I_G$ ratio increases, changing pH from acidic to basic values (see Table 2). Raman spectra for AA-functionalized GO are shown in Fig. 12. The ratio of D and G bands is an indicator of the functionalization level [44]. As seen from the spectra, there is a significant change in Raman shifts for the AA-functionalized GO. Table 2 and the spectra further support the successful functionalization of a GO-AA system.

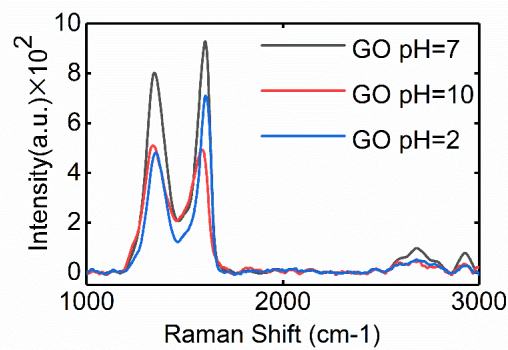

Fig. 11. Raman spectra of GO for acidic, basic, and neutral pH values.

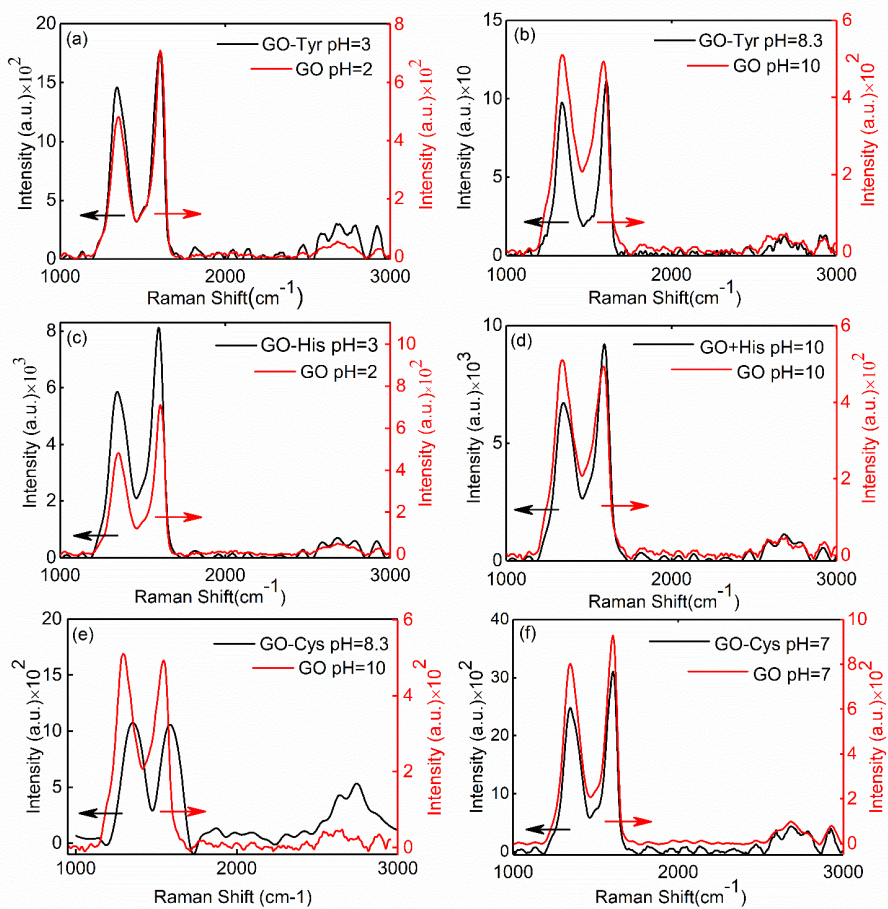

Fig. 12. Raman spectra of GO and (a and b) Tyr-, (c and d) His-, and (e and f) Cys-functionalized GO.

Based on Table 2, for AA-functionalized GO, the values of $I_D/I_G$ ratio at high (about 10) and low (about 3) pH values were almost similar and somewhat higher than that of GO, suggesting an increase in disorder in microstructures of functionalized GO. Besides D and G peaks, the peaks corresponding to the 2D, D + D', and 2D' bands of the second-order spectra are also pronounced, the intensity of which changes significantly depending on the pH and amino acid.

|  | Laser (nm) | pH | Raman shift (cm$^{-1}$) | Raman shift (cm$^{-1}$) | $I_D$ peak (counts) | $I_G$ peak (counts) | $I_D/I_G$ |
|---|---|---|---|---|---|---|---|
| **GO** |  | 2 | *1353* | *1607* | *482.036* | *709.819* | *0.679* |
|  |  | 7 | *1344.8* | *1604.5* | *801.709* | *928.226* | *0.863* |
|  |  | 10 | *1339.3* | *1588.5* | *508.271* | *493.047* | *1.03* |
| **rGO** |  | 7 | *1345.2* | *1603.3* | *1532.113* | *1672.447* | *0.916* |
| **GO+Tyr** | 532 | 3.03 | *1347.6* | *1585.8* | *71.601* | *63.062* | *1.135* |
|  |  | 8.3 | *1355.8* | *1583.2* | *1136.739* | *1076.412* | *1.056* |
| **GO+His** |  | 3.07 | *1596.5* | *1344.8* | *8119.272* | *5856.416* | *0.873* |
|  |  | 8.01 | *1344.8* | *1593.8* | *6711.664* | *9205.866* | *1.386* |
| **GO+Cys** |  | 7 | *1358.6* | *1591* | *2374.333* | *2852.766* | *0.832* |
|  |  | 8.06 | *1358.6* | *1388.7* | *1308.214* | *1075.454* | *1.216* |

Table 2. Analysis of Raman modes and their intensities.

For confirmation of GO-AA complex formation, fluorescence quenching was performed (Fig. 13), which shows a gradual decrease in Tyr absorption peak and redshift of GO absorption peak (the peak of photoluminescence for GO was obtained at 425 nm while after quenching, it was obtained at 480 nm). GO concentration increases from 1.0 mL to 1.6 mL for the 0.05 mg/mL GO/water solution.

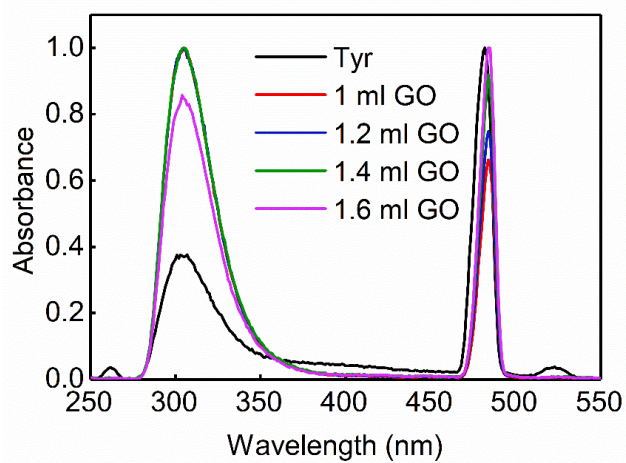

Fig. 13. Tyrosine PL quenching by GO.

**POM Investigation of GOLC**

Having a clear picture of GO-AA interaction mechanisms and knowing parameters (flake size, concentration, solvent medium, substrate treatment, and cell gap) that are crucial for the lyotropic LC phase formation of GO suspensions, we could observe a functionalized GOLC phase pH dependency by POM. Fig. 14 presents the optical micrographs of the L-His-GO system between crossed polarizers for two pH values. As seen from the figure, GOLCs show birefringence in ethanol to some extent; however, isotropic-to-LC phase transition concentration was found to be dependent on the solvent pH. Our experiments proved that the LC phase is obtained for pH above 8 compared to pH below 4 for the same concentration of GO flakes.

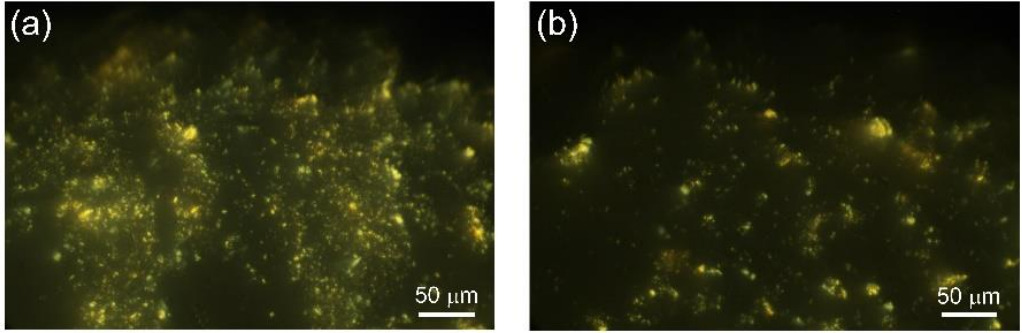

Fig. 14. POM images of His-functionalized GO

It is worth mentioning that for low concentrations of GO flakes in the biphasic state, ordered droplets of almost uniform size are observed (see Fig. 15a), which become highly organized and larger droplets after AA functionalization between nontreated glass substrates (see Fig. 15b), which requires further investigation.

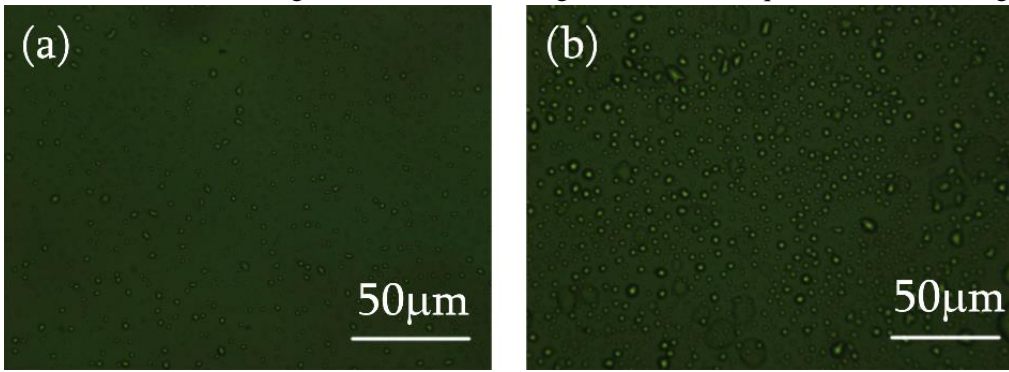

Fig. 15. POM micrographs of (a) GOLC cell between the crossed polarizers and (b) AA-functionalized GOLC cell between the crossed polarizers.

One of the most essential characteristics for the long-term operation of devices made based on LCs is the stability of the LC configuration structure, particularly the stability against various types of defects. Within the framework of this article, the effect of the magnetic field on the stability of the LC phase of the synthesized GO was studied. To manipulate the orientation of GO flakes, the external magnetic field was applied using an axially magnetized magnet (the magnetic flux intensity is about 500 mT). Microscope glass slides of 2 cm × 2 cm were washed in a sonication bath with acetone and methanol for 30 min. Sandwich cells with different substrate gaps of 10, 50, and 100 μm were prepared to study the LC properties of GO under the applied magnetic field using POM. Finally, the cells were glued with UV curable glue and capillary filled with the solutions of different GO-dispersed concentrations. The focal conic flowers formed in four directions between the crossed polarizers by the GOLC structure were observed because of the applied magnetic field (see Fig. 16). The observed structures correspond to the lamellar smectic phase.

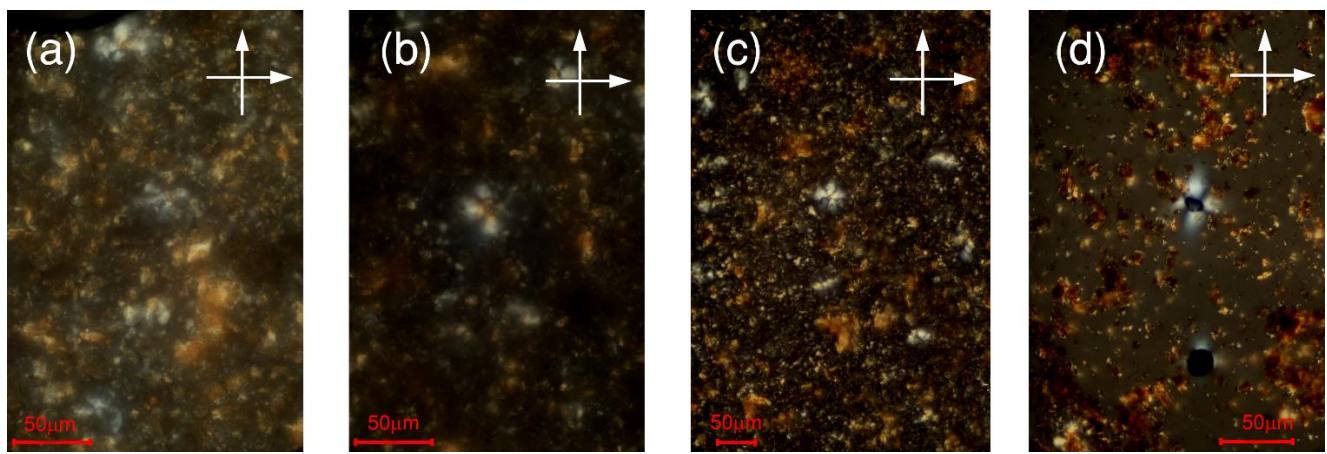

Fig.16. POM micrographs of GOLC between crossed polarizers operated in transmission mode: (a, b, c) seen with the whole cell immediately after removal of the magnetic field and (d) 1 weak after removal of the magnetic effect.

Further, the magneto-optical properties of GO and rGO LCs and the stabilization possibilities of the ordered alignment under other external stimuli, such as mechanical shearing, electrical field, and UV light, will be studied in detail [45, 46].

**Conclusion**

In summary, the functionalization of graphene oxide by amino acids, their reduction, and the formation of liquid crystalline phases were studied in this work. Based on the advanced Hummers' method, the oxidation efficiency was increased by controlling the flake size of GO *via* sonication, centrifugation, and filtration. The time evolution of synthesized materials was monitored over several months. They remained stable for the studied interval of time. Our spectroscopic studies proved that stable chemical bonds are formed due to the interaction between the observed biomolecules and graphene-like materials, which significantly changes their physical properties. In particular, the UV-Vis, Raman, and PL spectroscopy results revealed the interactions between three amino acids (tyrosine, histidine, and cysteine) and GO and rGO. Our UV-Vis spectrometric analysis showed $\pi-\pi^*$ stacking for AA-functionalized GO and blue and red shifts of absorption maxima in the case of different amino acids depending on the pH. The Raman spectra are distinguished by both pronounced characteristic maxima and peaks corresponding to the 2D, D + D', and 2D' bands of the second-order spectra, the intensity of which changes significantly depending on the pH. Complexation, redshift, and pH-dependent behavior of AA-functionalized GOLC were shown by the PL spectroscopy. XRD analysis also showed a peak shift due to the successful functionalization of GO and L-Cysteine dimer grew on GO. In addition, a paramagnetic behavior of GOLCs was observed. This is attributed to the presence of functionalized groups and other defects. Overall, our AA-functionalized GOLCs demonstrate pH-induced effects, which can be used in various applications, such as in biosensing, photothermal therapy in the biomedical field, and photoelectrochemical applications in nanocomposite membrane electrolytes for direct methanol fuel cells.

**Supplementary Material**

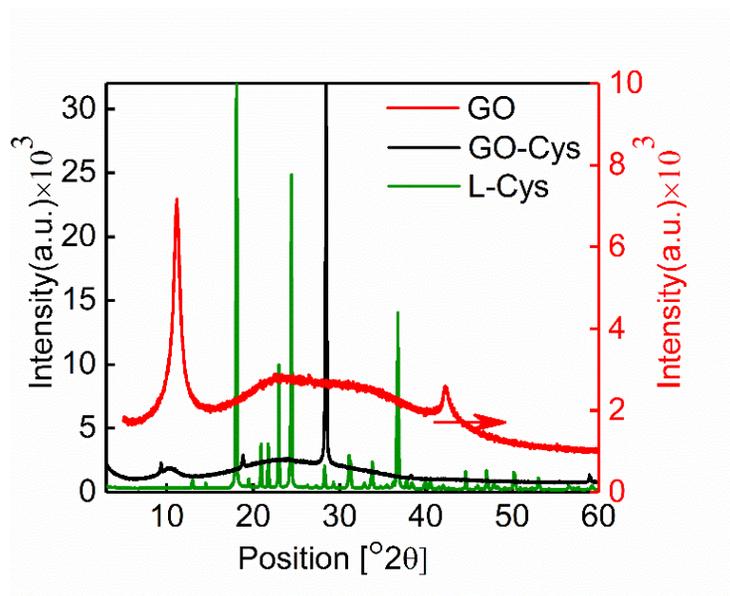

Fig. S1. XRD analysis of GO (red), L-Cys (green), and L-Cys functionalized GO (black).

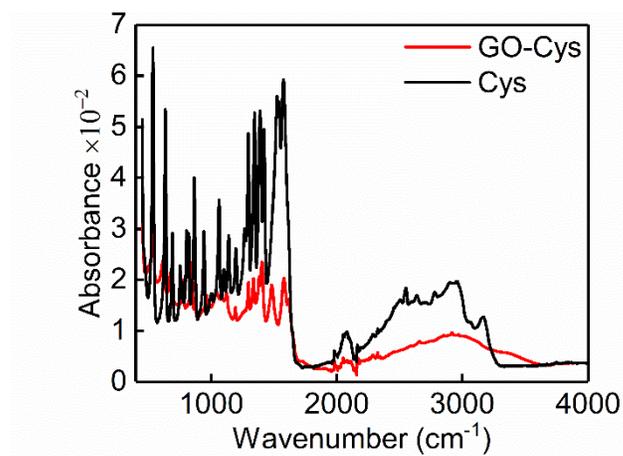

Fig. S2. FTIR-ATR spectra of L-Cys (black) and L-Cys functionalized GO (red).

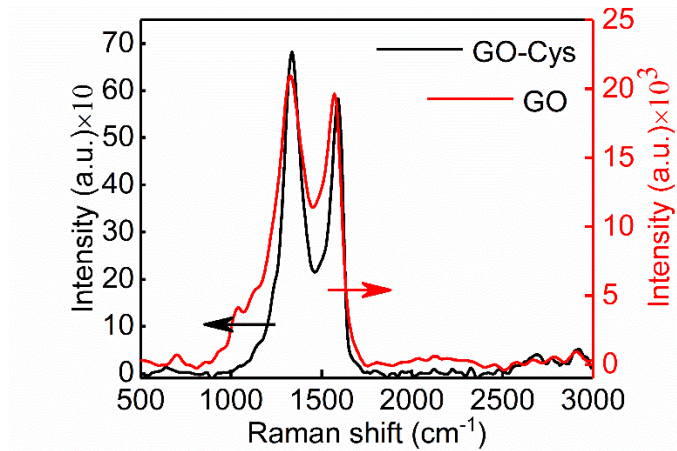

Fig. S3. Raman spectra of GO (red) and L-Cys functionalized GO (black).

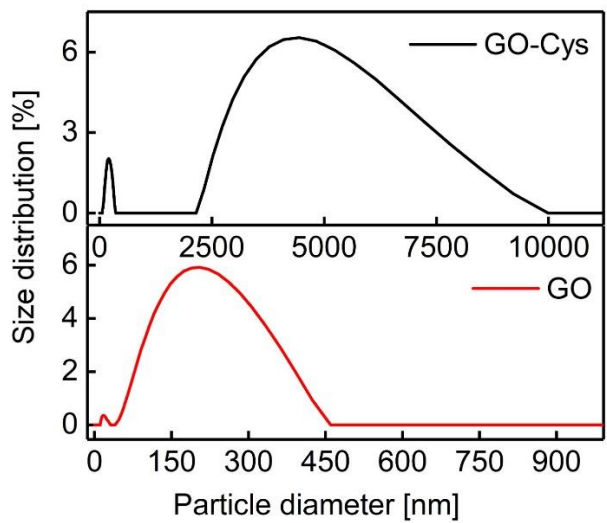

Fig. S4. Particle size analysis results of synthesized samples: GO (red) and L-Cys functionalized GO (black).